# Protein-ligand interaction study to identify potential dietary compounds binding at the active site of therapeutic target proteins of SARS-CoV-2

Seshu Vardhan, Bharat Z. Dholakiya and Suban K Sahoo[*]

*Department of Applied Chemistry, S.V. National Institute of Technology (SVNIT), Surat-395007, India. (E-mail: sks@chem.svnit.ac.in, suban_sahoo@rediffmail.com)*

**Abstract**

*Objective:* Total 186 biologically important phenylpropanoids and polyketides compounds from different Indian medicinal plants and dietary sources were screened to filter potential compounds that bind at the active site of the therapeutic target proteins of SARS-CoV-2.

*Method:* The molecular docking studies were carried out by using the Autodock Vina. The *in silico* ADMET and drug-likeness properties of the compounds were predicted from SwissADME server.

*Result:* The molecular docking study of the 186 compounds with the therapeutic target proteins (Mpro, PLpro, RdRp, SGp and ACE2) of SARS-CoV-2 resulted 40 compounds that bind at the active site with dock score above -8.0 kcal/mol.

*Conclusion:* Based on the *in silico* ADMET study and drug-likeness prediction of 40 compounds, we proposed petunidin, baicalein, cyanidin, 7-hydroxy-3',4'-methylenedioxyflavan, quercetin and ellagic acid among the 186 biologically important phenylpropanoids and polyketides as potential lead compounds, which can further be investigated pharmacologically and clinically to formulate therapeutic approaches for the COVID-19.

# 1. Introduction

The lack of recommended efficacious drugs or vaccines for the COVID-19 disease caused the ongoing pandemic, which affected 188 countries and territories with 5.20 million confirmed cases and death of 337,000 people by 22$^{nd}$ May 2020 [1]. The COVID-19 disease caused by the virus strain of severe acute respiratory syndrome coronavirus 2 (SARS-CoV-2) identified first in December 2019 at Wuhan, China. SARS-CoV-2, the single-stranded positive-sense RNA virus has *beta*-coronavirus genus with closely related genomic organization to SARS-CoV identified in 2003 [2]. The envelope of the spherical-shaped SARS-CoV-2 consists of structural proteins like membrane (M), envelope (E) and spike (S) proteins, where the club-shaped spike glycoprotein (SGp) interacts with the angiotensin-converting enzyme 2 (ACE2) of human cells and allows the SARS-CoV-2 virus to enter into the cells [3]. After entering cells, the viral replication and transcription is mediated with functional proteins like main protease (Mpro or 3CLpro), papain-like protease (PLpro), RNA-dependent RNA polymerase (RdRp) and helicase [4]. After the publication of the genomic sequence and crystal structures of various proteins of SARS-CoV-2 like Mpro, PLpro, SGp, RdRp etc. [5-8], research on structure-based molecular docking, dynamic simulations and quantum mechanical optimizations are performed to discover appropriate inhibitors for SARS-CoV-2 from the available databases/library containing approved drugs and phytochemicals with medicinal importance [9-10]. The computer based simulations and docking approaches are adopted to expedite the search to get target drugs/molecules from the library/databases that contained in lakhs, and also to minimize the experimental cost and time in developing appropriate drug/vaccine for COVID-19 [11].

The computational docking and simulations help in repurpose drugs like remdesivir, favipiravir and arbidol to fight against COVID-19 [12-14]. These repurpose drugs showed some promising results, but required further clinical evidences to examine their safety and efficacy in the treatment of COVID-19 [15-17]. Therefore, the computational docking and simulations research are ongoing globally to formulate new and effective therapeutic approaches for COVID-19. As a part of our ongoing research to search potential structures from phytochemicals that can effectively bind at the active site of various therapeutic target proteins of SARS-CoV-2 [18], we selected 186 phenylpropanoids and polyketides based phytochemicals for screening against the proteins 3CLpro, PLpro, SGp, RdRp and ACE2. The phenylpropanoids and polyketides are well known for diverse biological activities and pharmacological properties. Based on the *in silico* ADMET, drug-likeness and dock score, the compounds petunidin, baicalein, cyanidin, hydroxy-3',4'-methylenedioxyflavan, quercetin and ellagic acid among the 186 biologically important phenylpropanoids and polyketides were proposed as potential lead compounds, and their interactions with the target proteins were discussed.

## 2. Computational methods

The SDF files of the 186 biologically important phenylpropanoids and polyketides were retrieved from EMBL-EBI (www.ebi.ac.uk/chebi/advancedSearchFT.do) and PUBCHEM (https://pubchem.ncbi.nlm.nih.gov/). The structures were re-optimized by semi-empirical PM6 method by using the computational code Gaussian 09W and converted to PDB file format [19]. The crystallography protein structure of Mpro (PDB ID: 6LU7), PLpro (PDB ID: 4MM3), RdRp (PDB ID: 6M71), SGp-RDB (PDB ID: 2GHV) and ACE2 (PDB ID: 6M17) were downloaded from the PDB database (www.rcsb.org). The fine structure of the proteins were refined using Swiss model online tools followed by analysed using the Ramachandran plot. The ligand and protein structures were prepared for docking by using the MGL tool. The molecular docking studies were carried out by using the Autodock Vina1.1.2 [20]. The docked structures were analysed using the visualization tool BIOVIA Discovery studio. The important pharmacokinetic properties and ADMET (absorption, distribution, metabolism, excretion and toxicity) properties of the compounds were screened using the online tool SwissADME 'http://www.swissadme.ch/index.php' and 'http://biosig.unimelb.edu.au/pkcsm/prediction'.

## 3. Results and discussion

### 3.1. Screening of phytochemicals

Total 186 phenylpropanoids and polyketides found in the Indian medicinal plants and/or diets possesses diverse biological activities and pharmacological properties were collected from different databases, and screened against the five important therapeutic target proteins of SARS-CoV-2, i.e., Mpro, PLpro, RdRp, SGp and ACE2 by molecular docking analysis by using Autodock Vina. The dock score of 186 compounds with the five target proteins are summarized in **Table S1**. All the screened compounds showed good dock score, and we found that total 49, 24, 32, 63 and 77 compounds showed more than -8.0 kcal/mol dock score with each of the proteins Mpro, PLpro, SGp, RdRp and ACE2, respectively. To search the potential compounds, we performed protein-ligand interaction study that showed maximum binding affinity and dock score above -8.0 kcal/mol, and filtered the compounds that bound to the active site of any of the therapeutic target proteins of SARS-CoV-2. The protein-ligand interaction study revealed that only 40 compounds are binding at the active site of the therapeutic target proteins of SARS-CoV-2 (**Table S2**). These 40 compounds are considered for further screening by *in silico* ADMET study and drug-likeness prediction.

The ADMET/drug-likeness screening of 40 compounds resulted only 9 compounds (arjunolone, cyanidin, 7-hydroxy-3',4'-methylenedioxyflavan, isorhamnetin, 7,4'-dihydroxyflavone, quercetin, baicalein, petunidin and ellagic Acid) which obeying all the limitations of ADMET and drug-likeness properties (**Table S2**). The predicted drug-likeness properties along with the ADME parameters and pharmacokinetic properties from the SWISSADME and pKcsm online server are summarized in **Table S3**. The structure of these lead compounds can be studied for lead optimization to discover novel drugs for COVID-19. Also, these compounds can be used as an oral and intravenous admissible for further clinical trials. Therefore, a deeper study on the protein-ligand interaction study was performed to examine the modes of interaction of the lead compounds at the active site of the target proteins of SARS-CoV-2.

## 3.2. Protein-ligand interaction study

Based on the dock score and reported medicinal importance of the 9 potential compounds (**Table S3**), 6 lead compounds were proposed and their mode of interactions with the residues of target proteins of SARS-CoV-2 at the active site are presented in Table 1. Compared to our recent proposed triterpenoids based lead compounds i.e., 7-deacetyl-7-benzoylgedunin (-9.1 kcal/mol), epoxyazadiradione (-8.4 kcal/mol), limonin (-9.2 kcal/mol), maslinic acid (-9.3 kcal/mol) and glycyrrhizic acid (-10.3 kcal/mol) respectively for the target protein Mpro, PLpro, RdRp, SGp and ACE2, the phenylpropanoids and polyketides based lead compounds showed slightly lower dock score, except for PLpro [18]. However, their distinct structure from the triterpenoids with the different mode of interactions at the active site of SARS-CoV-2 makes them attractive inhibitors. Also, these lead compounds are mainly available in dietary sources and possesses diverse pharmacological properties makes them ideal choice for formulating new therapeutic approach for COVID-19.

**Table 1.** The interactions of the proposed lead compounds with the target proteins of SARS-CoV-2.

| Sources | Lead Compounds | Target Protein | B.E (Kcal/mol) | Interactions |
|---|---|---|---|---|
| Vitis vinifera | Petunidin | Mpro | -8.2 | Chain A: Hydrogen Bond: SerA144, Gly143, His163. Carbon Bond: Cys145, π -alkyl: Met165, π - π: His41, Pi-Donor: Phe140. VDW: Asn142, Leu141, His172, Glu166, Leu167, Pro168, Gln189, Arg188, His164, Met49, Asp187. |
| Oroxylum indicum | Baicalein | Mpro | -8.1 | Chain A: Hydrogen bond Thr6, Thr24, unfavourable donor: Cys145, Pi-cation: His41, Carbon bond: Thr25, VDW: His163, Met165, Asn142, Leu141, Ser144, Met49, Gly143, |

| | | | | Ser46, Thr45, Leu27, Glu166, Gln189. |
|---|---|---|---|---|
| Vitis vinifera | Cyanidin | PLpro | -8.5 | Chain B: Hydrogen bond: Gly267, Tyr274, Asp165. π - π: Asn268, Tyr265, Carbon bond: Pro249. VDW: Thr302, Gly164, Tyr269. |
| Terminalia bellirica | 7-Hydroxy-3',4'-methylenedioxyflavan | SGp-RBD | -8.2 | Chain C: Hydrogen Bond: Arg444, His445, Pi-Pi- Phe460, VDW: Asp454, Leu443, Pro477, Arg441, Val456, Pro459. Chain E: Hydrogen bond: Pro459, Cys467, π -cation: His445, π -alkyl-Pro466, Val458, VDW: Phe 460, Pro477, Leu443, Arg441, Arg444, Ser456, Asp454. |
| Allium cepa | Quercetin | RdRp | -9.0 | Chain A: Hydrogen Bond: Arg349, Val315, Asn628. π –alkyl: Pro677, Arg349, Val315, Pro461, π –sigma: Pro461, π –cation: Arg349, VDW: Val675, Phe396, Cys395, Arg457, Thr319, Ser318. |
| Punica granatum | Ellagic acid | ACE2 | -8.4 | Chain B: Hydrogen bond: Gln98, Asn210, Trp566, Asp206, π –alkyl: Leu95, Val209, VDW: Val212, Leu91, Pro565, Glu208, Ala396, Asn397, Lys562. |

### 3.2.1. Mpro or 3CLpro (Chymotrypsin like protease)

The main protease cleaves about 11 polyproteins including its N and C terminal auto processing sites for the generation of the individual protein units that includes RdRp, helicase, ssRNA-binding protein, endonuclease, exoribonuclease and 2-O-ribise methyltransferacse etc. The target protein Mpro play important role in replication and transcription of SARS-CoV-2 virus. Mpro contains two domains, i.e., domain I (chymotrypsin) and domain II (picornavirus 3 C protease-like). Each domain contains six-stranded antiparallel β-barrel structure and the catalytic site is present at the residues Cys145 and His41 [21-23]. Among the screened 9 compounds, the petunidin and baicalein binding at the active site of Mpro with the dock score -8.2 and -8.1 kcal/mol, respectively (**Table 1**).

Petunidin (anthocyanidin cation) is a phenolic compound and a plant metabolite which is found in grapes and red berries constituting 5-hydroxyl substituent's and 7-methoxy substituent's. It is identified as a potential inhibitor of Mpro. The posing ligand exclusively binding to the active site of the target Chain A. Some of the pocket residues interactions showing hydrogen Bonds to SerA144, Gly143, His163, which indicates the strong binding affinity. Petunidin rings effectively maintaining carbon bond to Cys145 and π-alkyl bond to Met165. The target protein catalytic dyed residues posing π-π bond to His41 and π-donor bond to Phe140 interactions. The closest VDW non-covalent interaction to Asn142, Leu141, His172, Glu166, Leu167, Pro168, Gln189, Arg188, His164, Met49 and Asp187 residues allows to fit ligand in the active pocket (**Fig. 1**). The other compound baicalein also showed nearly same binding affinity to Chain A residues by forming hydrogen bonds to Thr6, Thr24, π-cation complex to His41 and unfavourable donor to Cys145 catalytic dyed residues (**Fig. S1**). Other residues of target protein participate in forming non-covalent interactions are summarized in Table 1. The petunidin and other anthocyanins synergistically shows carcinogenic effect. In compared to petunidin, the medicinal importance of baicalein is well known [24]. Baicalein shows anti-inflammatory and antitumor effects. It is used in Chinese traditional medicine for treating liver disorders. Also, baicalein acts as an anti-furin activity in CT-26 cells proliferation and migration resulting anti-cancerous properties [25], which makes baicalein an important inhibitor as the SARS-CoV-2 spike glycoprotein contains a furin cleavage site [26].

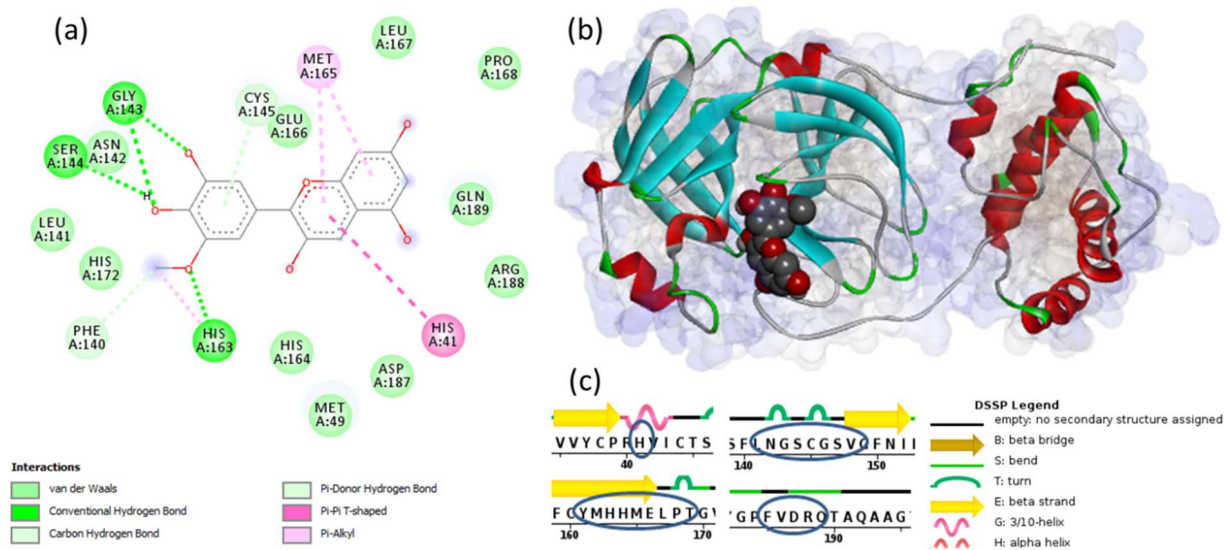

**Fig. 1.** (a) 2D representation showing various non-covalent interactions between petunidin and Mpro, (b) 3D view showing the position of petunidin within the cavity of Mpro, and (c) sequence indicating potions of interactions.

### 3.2.2. PLpro (papain like protease)

PLpro, a cysteine protease that mediates SARS-CoV-2 RNA replication [27]. It functionally serves as a cleaving machine spanning nsp1 through nsp4. It is also involving the deubiquitinating enzyme in aggressive proliferation of disease. PLpro consists of hydrophobic cavities and is highly conserved to substrate binding profiles especially S2/S4 domains. PLpro efficiently cleaves the replicase substrate with the function of catalytic triad (Cys112-His273-Asp287) and zinc-binding site. This pocket plays a key role to find the potent inhibitor for PLpro, and identified the potent lead compound that could stop the RNA replication of SARS-CoV-2.

Among the screened compounds having drug-likeness properties, the phenolic compounds like arjunolone, cyanidin, 7-Hydroxy-3',4'-methylenedioxyflavan, Isorhamnetin, and 7,4'-Dihydroxyflavone bind at the active sites of PLpro with dock score -8.3, -8.5, -8.3, -8.3 and -8.4 kcal/mol, respectively. These compounds are medicinally reported for their antioxidant, anti-inflammatory and antiviral properties. Most of them are found in edible fruits that collectively known for nutraceutical compounds. The modes of interactions of the cyanidin showing higher dock score with the PLpro are summarized in **Table 1**.

Cyanidin is a pigment found in berries enhancing red, orange and blue colours. Cyanidin is a polyketide and anthocyanidin pigment found in grapes and red berries. Cyanidin and its glycoside analogue pose medicinal values such as anti-inflammatory, anti-diabetes, anti-obesity and vasoprotective [28]. Anthocyanins structurally similar to cyanidin show anti-cancerous properties and lower the risk of cancer and heart diseases. Cyanidin actively binding to the pocket of catalytic triad residues of PLpro. It is showing binding affinity to PLpro and can be served as an inhibitor. Cyanidin binding to active site of PLpro by forming a complex with Chain B residues showing hydrogen bonds to Gly267, Tyr274 and Asp165 residues. The active site is a hydrophobic cavity showing the ligand forming π- π interactions to Asn268, Tyr265 residues and carbon bond to Pro249 (**Fig. 2**).

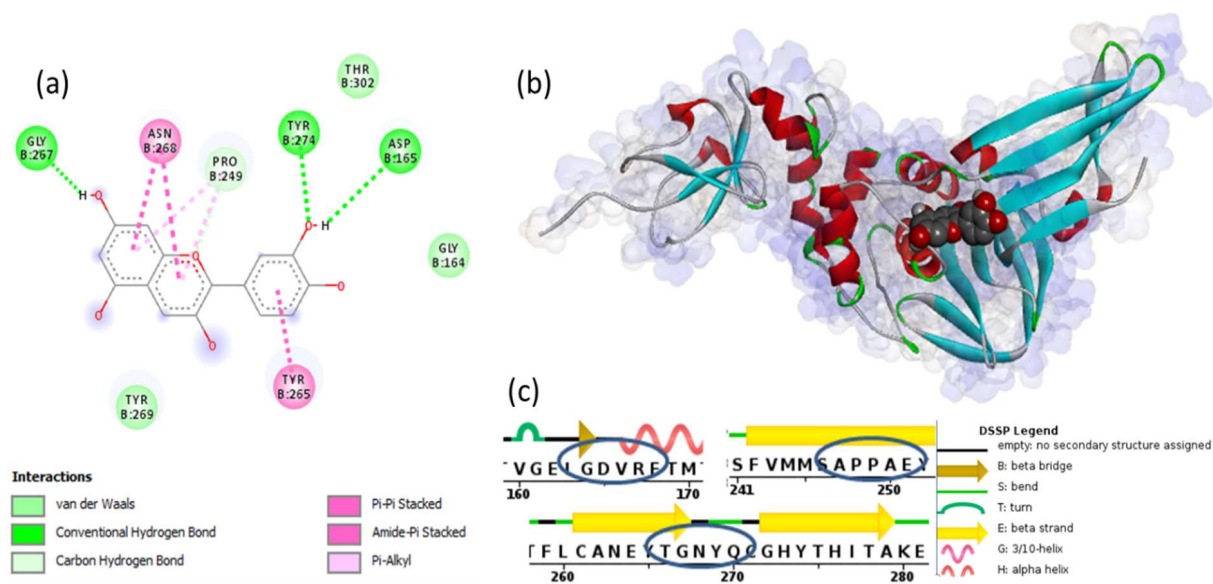

**Fig. 2.** (a) 2D representation showing various non-covalent interactions between cyanidin and PLpro, (b) 3D view showing the position of cyanidin within the cavity of PLpro, and (c) sequence indicating potions of interactions.

### 3.2.3. SGp-RBD (Spike glycoprotein-receptor binding domain)

SGp plays a vital role in SARS-CoV-2 disease establishment by attaching the cell membrane of the human receptor ACE2 and TMPRSS2 [29,30]. This also internalizes the virus into the endosomes where the conformational changes take place in the S glycoprotein. This protein promotes the fusion between the viral protein and the cell membrane constituting 3 states namely pre-fusion native, pre-hairpin intermediate state and post-fusion state. It acts as a viral fusion peptide that covers up S2 cleavage that occurs during virus endocytosis. These spike proteins determine the virion-host tropism that includes the entry of the virions into host cells and their interactions with human ACE2. The exposed portion of SGp consists of RBD, where substrate domain binding takes place. The fusion peptide is a targetable domain to develop the therapeutic compounds. We had screened phytochemicals through molecular docking to identify potent inhibitors for SGp that could practically obstruct the SGp-ACE2 and SGp-TMPRSS2 complex formation and activation. The SGp-RBD consists hydrophobic sites and substrate binding cavities between 333-527 amino acid residues of chain C and chain E.

The compounds 7-hydroxy-3',4'-methylenedioxyflavan binding to the active sites of side chains C and E of SGp-RBD, and forming inhibitory complex with dock score -8.2 kcal/mol (**Fig. 3**). It is forming a complex through hydrogen bonds with Arg444, His445 residues, π-π interaction with Phe460 and have non-covalent VDW forces with Asp454, Leu443, Pro477, Arg441, Val456 and Pro459 residues (**Table 1**). Also, 7-hydroxy-3',4'-methylenedioxyflavan binding to Chain E amino acid residues forming hydrogen bonds with Pro459, Cys467, π-cation interaction with His445, and π-alkyl bond with Pro466, Val458 (**Fig. S2**). The nearest non-covalent VDW interactions with Phe460, Pro477, Leu443, Arg441, Arg444, Ser456, Asp454. Medicinally, 7-hydroxy-3',4'-methylenedioxyflavan is an antiviral drug that is reported for activity against the production of HIV1 antigen in MT4 cells and regression of cell proliferation of MCF7 of BUS cells [31].

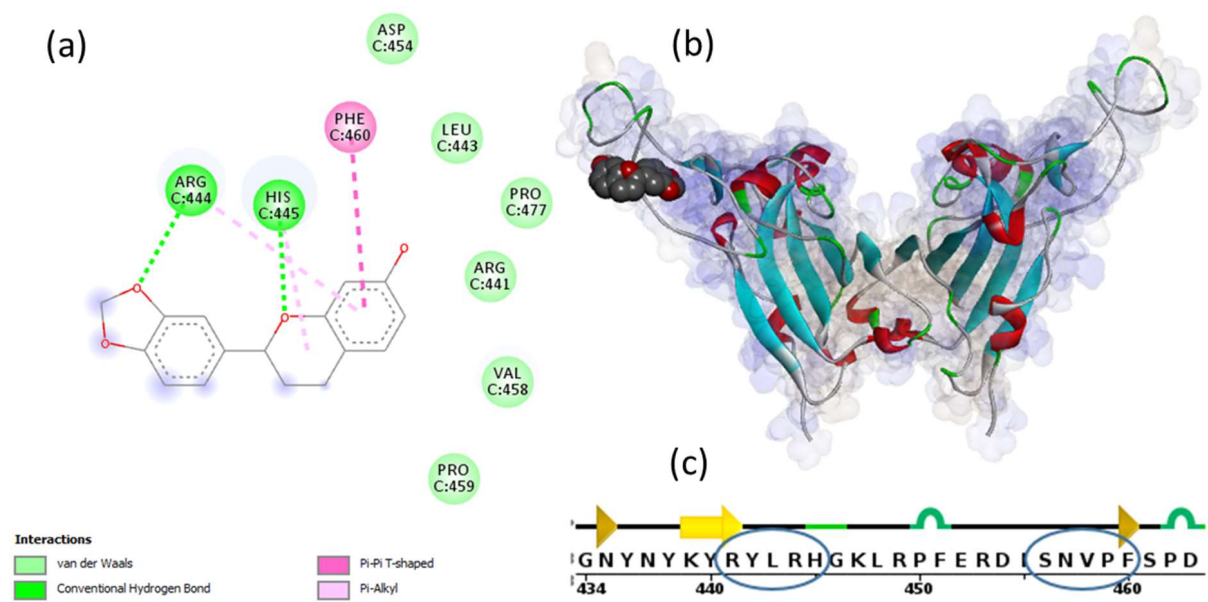

**Fig. 3.** (a) 2D representation showing various non-covalent interactions between 7-hydroxy-3',4'-methylenedioxyflavan and SGp-RBD, (b) 3D view showing the position of 7-hydroxy-3',4'-methylenedioxyflavan within the cavity of SGp-RBD, and (c) sequence indicating potions of interactions.

### 3.2.4. RdRp (RNA-Dependent RNA polymerase)

RNA-Dependent RNA polymerase (RdRp) is well known as nsp12, which plays major role in the transcription and replication process of SARS-CoV-2 [32]. Recently identified structural changes in RdRp and functional residues are involved in replication. Nsp12 consists of beta haipin like domain in N terminus. It is structurally similar to SARS-CoV RdRp, having nsp12-nsp7-nsp8 complex proteins and functional domains in replication. In nsp12 residues from 4-118 constitutes an antiparallel β-strands and two helices. Residues from 215-218 form a β-strand of nsp12 domain. The NiRAN domain and the palm subdomain in the RdRp domain forms a set of close contacts to stabilize the overall structure. The polymerase domain a fingers subdomain residues are 366 to 581 and 621 to 679, a palm subdomain 582 to 620 and 680 to 815, and a thumb subdomain residues 816 to 920. The active site of the SARS-CoV-2 residues from 753 to 767 that contains the catalytic residues from 759 to 761, in which the serine and glycine amino acids shows most catalytic functions during polymerization and transcription. The active site poses a hydrophobic cavity between nsp12 and nsp7 residues.

Among the 9 lead compounds, computational study revealed quercetin as potent inhibitor of RdRp and binding to polymerase site chain A posing hydrogen bonds to Arg349, Val315, Asn628 and showing π-alkyl, π-cation and π-sigma interactions to Pro677, Arg349, Val315, Pro461, Val675 residues (**Fig. 4**). Quercetin forming a strong non-covalent VDW interactions with Phe396, Cys395, Arg457, Thr319, Ser318 (**Table 1**). In the site2 quercetin binding to the Gln573 and Ser682 residues forming hydrogen bond. The binding of these residues including glycine and serine amino acids of active site cavity with quercetin could hinder the function of RdRp in polymerization. The β strand of nsp12 domain is also encapsulated with quercetin at the active sites 1 and 2.

Quercetin, a flavonoid found in onions, berries and apples etc that is biologically active antioxidant and inflammatory compound. It is an orally admissible drug under investigation that prevents haemolytic, cardiovascular diseases and some of the chronic disorders like diabetes, arthritis, cancer etc. [33].

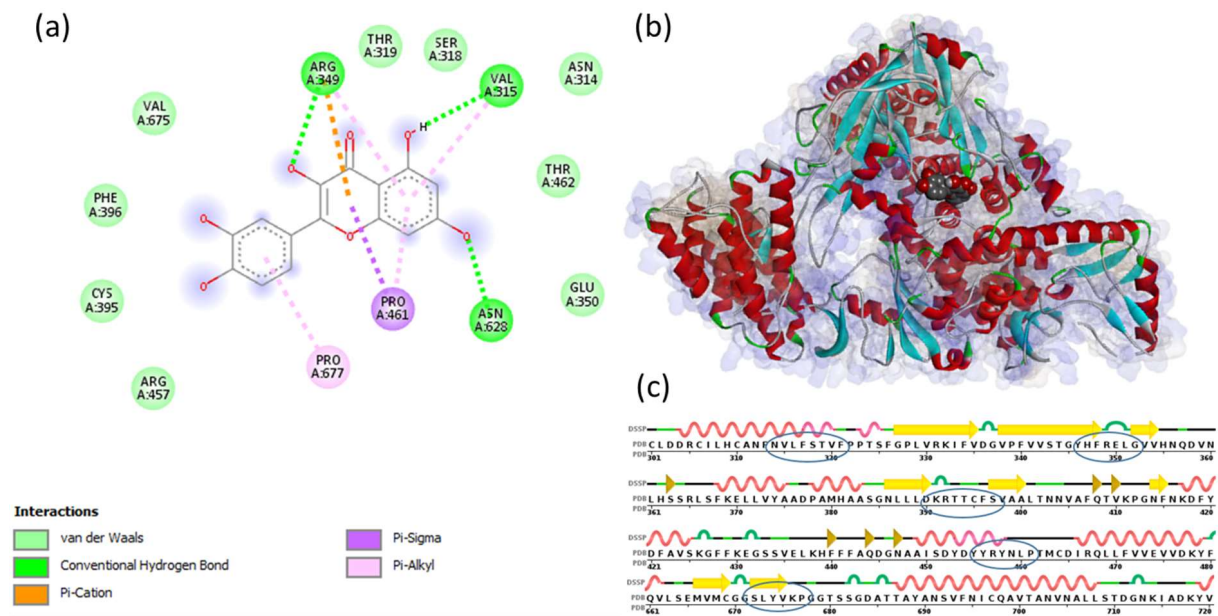

**Fig. 4.** (a) 2D representation showing various non-covalent interactions between quercetin and RdRp, (b) 3D view showing the position of quercetin within the cavity of RdRp, and (c) sequence indicating potions of interactions.

### 3.2.5. ACE2 (angiotensin-converting enzyme 2)

ACE2 is a zinc associated metalloenzyme and carboxypeptidase which is located on the endothelial cells of different organs especially arteries, lungs, kidneys etc. [34]. It is a type 1 membrane protein where its enzymatically active domain is manifested on the surface of the cells of lungs and other tissues. The interactions of the spike glycoprotein of the coronavirus with this protein results in the endocytosis and translocation of enzyme and the virus, which in turn act as entry point for the SARS-CoV-2 virus. The extra structural domains of the ACE2 are classified into two sub domains that are being connected at the base and the hinge bending movement that opens up the active site. The site directed mutagenesis is used to study the role of different active site residues. Arg273 plays a key role in the process of substrate binding where as its replacement results in the abolishment of the enzymatic activity. However, the His505 and His345 are entailed in the process of catalysis,

His345 acts as a hydrogen bond acceptor or receptor resulting in the formation of the tetrahedral peptide intermediate. ACE2 recognizes RBD with the protease domain of ACE2 mainly engages the $\alpha_1$-helix partially from the $\alpha_2$-helix and the linker of the $\beta_3$ and $\beta_4$ sheets.

Among the 9 lead compounds, the compounds ellagic acid (**Fig. 5**) and 7-hydroxy-3',4'-methylenedioxyflavan (**Fig. S3**) binds at the active site of ACE2 with the same dock score of -8.4kcal/mol. The types of interactions between the ellagic acid and ACE2 are summarized in Table 1. The ligand ellagic acid binding at chain B residues forming hydrogen bonds to Gln98, Asn210, Trp566, Asp206, $\pi$–alkyl:, VDW: Val212, Leu91, Pro565, Glu208, Ala396, Asn397, Lys562. It is also showing $\pi$-alkyl interactions Leu95 and Val209 residues respectively. Ellagic acid is partially binding to the substrate binding pocket, the active site residues lies between 21-95, 335-400 and 501-527 amino acid residues. Ellagic acid is richly found in raspberries, blackberries, strawberries and other fruits and nuts. It is widely used as potent drug to treat cancers, bacterial and viral diseases [35].

Isorhamnetin is a polyphenolic compound found in fruits, vegetables and medicinal plants. It is showing binding affinity to active site of ACE2 substrate binding domain with partial contact. The ligand isorhamnetin posing comparably lower binding energy -7.6kcal/mol with chain B residues by forming hydrogen bonds to Gln98, $\pi$–alkyl to Leu95, Val209 and non-covalent VDW to Val212, Pro565, Glu208, Ala396, Gly205, Lys562, Tyr196, Gln102, Ala99, Trp566, Asn210 (**Fig. S4**). Another compound hesperidin also a phenolic compounds showing similar binding characteristics to ellagic acid, and binds to Leu95, Gln98, Gln102, Asn194, Asn210, Arg219, Ala396 and Lys562 residues with binding energy -9.2kcal/mol, but this compound violating Lipinski rule for drug-likeness properties. Therefore, the compounds ellagic acid and 7-hydroxy-3',4'-methylenedioxyflavan were proposed as potent compounds that can inhibit the activity of ACE2.

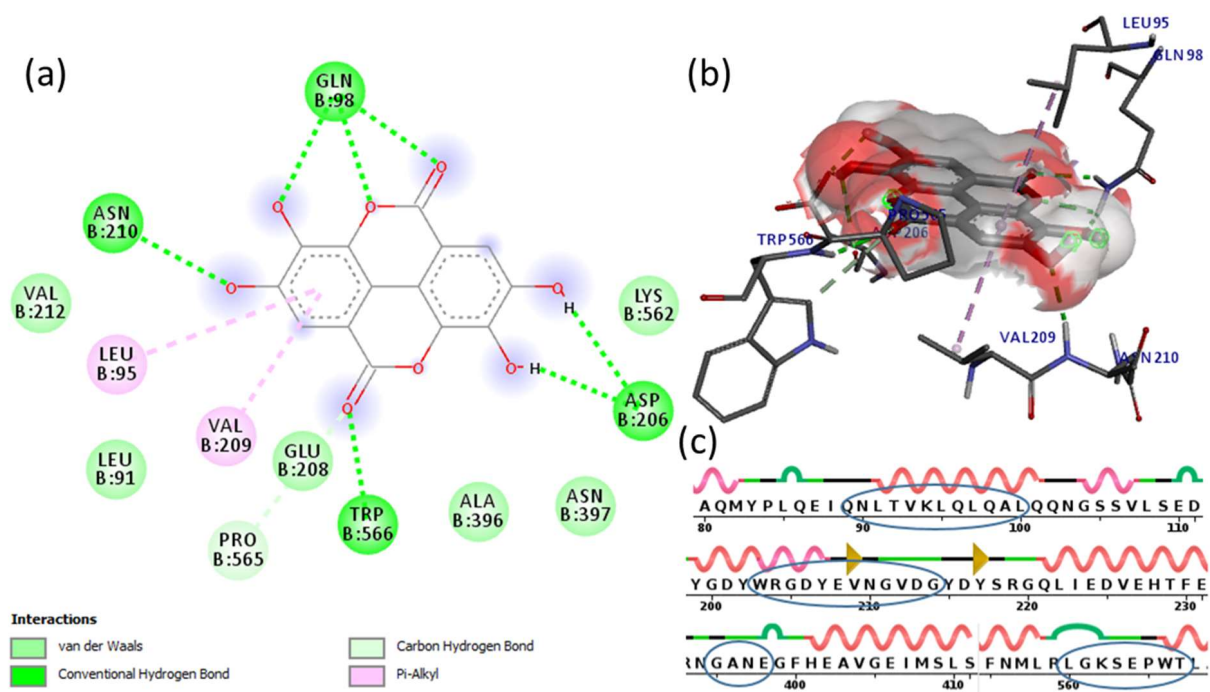

**Fig. 5.** (a) 2D representation showing various non-covalent interactions between ellagic acid and ACE2, (b) 3D view showing the position of ellagic acid within the cavity of ACE2, and (c) sequence indicating potions of interactions.

## 4. Conclusions

In summary, from the *in silico* computational screening of 186 phenylpropanoids and polyketides based phytochemicals found mainly in Indian medicinal plants and dietary sources, 6 lead compounds (petunidin, baicalein, cyanidin, 7-hydroxy-3',4'-methylenedioxyflavan, quercetin and ellagic acid) were proposed as potential against the therapeutic target proteins of SARS-CoV-2. Among the 6 lead compounds, the petunidin and baicalein was proposed to inhibit the function of Mpro. The compounds cyanidin, 7-hydroxy-3',4'-methylenedioxyflavan, quercetin and ellagic acid were proposed for the target proteins PLpro, SGp-RBD, RdRp and ACE2, respectively. These lead compounds are dietary and known for diverse medicinal properties. Therefore, formulation of therapeutic approach for COVID-19 will be easier provided they found experimentally effective in inhibiting the SARS-CoV-2. In addition, it is important to mention here that whenever the phytochemicals are examined, the novel drug delivery systems should be adopted for effective release of the phytochemicals at the target sites. Finally, the structure of the lead compounds can be studied further for lead optimization to discovery novel drug for COVID-19.

**Declaration of competing interest**

The authors declare that they have no known competing financial interests or personal relationships that could have appeared to influence the work reported in this paper.

**Acknowledgments**